# High-Efficiency Electrically Switchable Nonvolatile Thermal Transistor with Multiple Thermal Conductivity States Based on Ferroelectric HfO$_2$

Yong-Kun Huo, Hui-Feng Feng, Chao Yao, Zhong-Xiao Song, Zhi-Xin Guo*

*State Key Laboratory for Mechanical Behavior of Materials, Xi'an Jiaotong University, Xi'an, Shaanxi, 710049, China.*

**ABSTRACT:** While nanoscale electronic logic circuits are well-established, the development of nanoscale thermal logic circuits has been slow, mainly due to the absence of efficient and controllable nonvolatile field-effect thermal transistors.

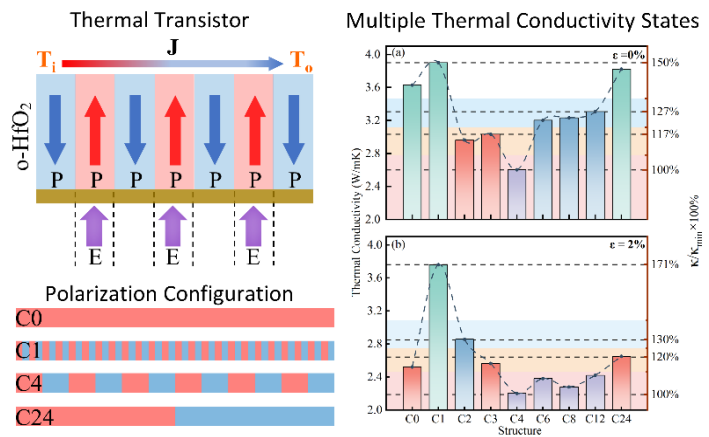

In this study, we introduce a novel approach that leverages ferroelectric orthorhombic hafnium dioxide (o-HfO$_2$) thin films to achieve electrically switchable nonvolatile field-effect thermal transistors. Using molecular dynamics simulations and machine learning potentials, we demonstrate that a 24 nm o-HfO$_2$ film can exhibit four distinct, reversible states of thermal conductivity. Notably, these states achieve a maximum switching ratio of 171% under 2% tensile strain. Our results underscore the potential of ferroelectric materials, particularly o-HfO$_2$, in advancing thermal logic circuits by enabling multiple, stable thermal conductivity states controlled by electric fields.

Heat energy, mainly arising from vibrations within a material's lattice structure—known as phonons—is often considered waste in electronic communication and detrimental to information processing. However, in certain situations, thermal signals can outperform electrical ones[1], suggesting that controlling heat could unlock novel and unexpected applications. Efforts to utilize heat flow in technologies similar to electronic circuits began in 1994, when Koller and colleagues introduced the concept of thermal logic circuits[2]. This initiative sparked interest in using phonons for transmitting and processing information. Yet research on thermal transistors and logic circuits progressed slowly until 2007, when Li and collaborators proposed a detailed model of a thermal transistor based on negative differential thermal resistance[3]. They also envisioned thermal logic gates capable of basic logical operations. Despite these theoretical advancements, realizing practical thermal transistors with controllable and efficient performance remains an ongoing challenge. This is especially true for nonvolatile field-effect thermal transistors, which, despite progress in nonvolatile field-effect electric transistors, are still underexplored.

Ferroelectric materials, characterized by their spontaneous polarization and the ability to reverse polarization under an external electric field, are ideal candidates for developing controllable thermal transistors. The thermal conductivity (TC) of these materials is strongly influenced by their polarization structure, making them highly suitable for thermal logic circuits. For instance, the perovskite oxide $PbTiO_3$ exhibits polarization-dependent anisotropic TC, with a high-to-low TC ratio reaching up to 200% at room temperature[4]. However, traditional perovskite oxides are limited to only two or three distinct TC states and suffer from depolarization effects at the nanoscale[5], significantly restricting their practical application and flexibility in thermal transistor design.

In contrast, the orthorhombic ferroelectric phase of $HfO_2$ (o-$HfO_2$, space group $Pca2_1$)[6] stands out due to its exceptional controllability. It enables polarization switching within domains as small as 2.7 Å[7] and achieves switching times on the nanosecond scale[8], while maintaining stable polarization states even after the removal of an external electric field. This nonvolatility, coupled with its ability to support multiple distinct TC states controlled by localized electric fields, positions o-$HfO_2$ as a highly versatile and scalable material[9-15]. These characteristics make it far superior to traditional ferroelectric materials in terms of controllability and operational flexibility, establishing it as a strong contender for nonvolatile, electrically controlled thermal logic circuits.

This study presents an electrically controlled thermal transistor based on o-$HfO_2$, showcasing remarkable performance and precise control. Through machine learning potentials and molecular dynamics simulations, we demonstrate that a 24 nm-long o-$HfO_2$ film can achieve four distinct and switchable TC states with switching ratios exceeding 10%, reaching up to 150% at room temperature. Notably, applying a 2% tensile strain further enhances the switching ratio to 171%. The underlying mechanism driving the polarization-dependent thermal conductivity in o-$HfO_2$ is elucidated through detailed analyses of phonon transmission spectra, phonon dispersion relations, density of states, and phonon participation rates. This work positions o-$HfO_2$ as a leading platform for high-efficiency, flexible, and electrically tunable thermal transistors, paving the way for next-generation thermal management technologies.

## ■ RESULTS AND DISCUSSION

A schematic diagram of the thermal transistor based on o-$HfO_2$ is shown in Fig. 1(a). In this design, o-$HfO_2$ serves as the core material, supported by a dielectric layer (depicted in brown). By applying a localized electric field to o-$HfO_2$, polarization switching can be induced, enabling multiple TC states. This mechanism provides precise control over the relationship between input and output temperatures.

The thermal switching performance of the o-$HfO_2$ thermal transistor is closely linked to its polarization configurations, necessitating a detailed understanding of how polari-zation states influence thermal transport. Achieving multiple stable and switchable polarization states in a single system relies on two factors: designing fundamental structural units and arranging them periodically to create desired configurations. Figure 1(b) illustrates the unit cell structure of o-HfO$_2$ in both upward and downward polarized states, differentiated primarily by the position of oxygen atoms.

To create multiple stable polarization states, external stimuli are employed to locally modulate polarization. Unlike perovskite ferroelectrics such as PbTiO$_3$, the intrinsic dipole splitting in o-HfO$_2$ plays a critical role in its domain structure. In uniformly polarized o-HfO$_2$, alternating layers exhibit zero-thickness domain walls (180° domain walls), enabling localized polarization control[7]. By applying a finely tuned electric field, polarization within a narrow region (as small as a single unit cell) can be reversibly switched. This ability to precisely control the polarization arrangement allows for tuning TC and tailoring thermal transport properties for thermal logic circuits and device applications.

To investigate how polarization configurations impact the TC of o-HfO$_2$, molecular dynamics simulations were conducted on structures with fixed dimensions: 24 nm in length (x direction), 2 nm in thickness (y direction, with periodic boundary conditions), and 2 nm in width (z direction). The primary variable is the periodic length $l$ along the x direction, as shown in Fig. 1(d). For example, the C1 configuration alternates single layers of up-polarized and down-polarized cells, while C2 alternates two layers of each. These configurations are designed to capture a broad range of TC behaviors.

This study calculates the TC of o-HfO$_2$ using the non-equilibrium molecular dynamics (NEMD) method, employing a deep neural network-based interatomic force field developed by Liu et al.[16] (see Supplemental Material for details). Figure 2 illustrates the calculated TCs for various polarization configurations. By analyzing these results, we explore the relationship between o-HfO$_2$'s TC and its polarization configurations, which is crucial for optimizing ferroelectric materials in thermal logic circuits. Under unstrained conditions, the TC of the C0 structure (with all polarization directions pointing upward) is 3.6 W/mK, which is lower than the 14.1 W/mK predicted by Zhang et al.[17] through the first-principle calculations. This discrepancy is mainly due to the small size (24 nm in length) of the o-HfO$_2$ structure in our study, exhibiting a clear size effect[18]. Interestingly, the TC of the standard antiferroelectric (AFE) structure C1 is higher than that of C0, reaching 3.9 W/mK. This result is unusual, as the AFE structure typically has more phonon scattering interfaces between different polarization domains, which would usually reduce TC. As discussed later, this phenomenon is attributed to the unique phonon band structure in the AFE configuration.

Figure 2 also reveals a strong polarization dependence of TC in o-HfO$_2$, which is promising for thermal transistor applications. As the periodic length $l$ increases, TC initially decreases but then rises, showing a non-trivial periodic-length dependence. Specifically, the C4 structure exhibits the lowest TC (2.6 W/mK) due to its polarization arrangement, with a switching ratio of up to 150% compared to C0 under unstrained conditions. These configurations can be classified into four stable TC states with a discrepancy greater than 10%, as shown in Fig. 2(a). This result demonstrates that four distinguishable TC states can be realized in a 24 nm o-HfO$_2$ structure, achieved by modulating the ferroelectric polarization states. This offers new possibilities for efficient thermal management in thermal logic circuits.

To investigate the maximum TC switching ratio achievable by different polarization configurations, we further examine the effect of strain on TC. Previous studies have shown that o-HfO$_2$ retains its *Pca2$_1$* space group characteristics under 2% and 2.5% tensile strain, maintaining ferroelectric properties within the polar *mm2* point group. However, at 3% tensile strain, the structure transitions to the non-polar *Pbcn* space group, resulting in a loss of ferroelectricity[19]. Therefore, we focus on the 2% tensile strain configuration to study its effect on TC. As shown in Figure 2, applying 2% tensile strain causes a decrease in TC for all structures, with the extent of reduction varying depending on the polarization configuration. Notably, under strain, the TC decreases to 3.8 W/mK for the C1 structure and 2.2 W/mK for the C4 structure, marking the highest and lowest TC values, respectively. The tensile strain reduces TC but expands the range of tunable TC, increasing the switching ratio between C1 and C4 to 171%. This strain-enhanced effect demonstrates that applying mechanical strain can broaden the tunable range of o-HfO$_2$'s TC, enabling multi-state reversible thermal switches and complex thermal logic operations.

Based on the TC variations under 2% tensile strain, we can classify the TC of different polarization configurations of o-HfO$_2$ into four distinct states: 2.2–2.4 W/mK, 2.5–2.6 W/mK, 2.9 W/mK, and 3.8 W/mK. Since each polarization configuration can be reversibly switched by adjusting the local electric field, these distinguishable TC states lay the groundwork for implementing multi-valued logic[20]. The ability to switch between multiple TC states provides a significant advantage over traditional binary systems, making ferroelectric o-HfO$_2$ a promising candidate for neuromorphic computing applications.

To uncover the mechanism behind the strong polarization-dependent TC in o-HfO$_2$, we calculated the phonon transmission spectra (PTS)[21] for several representative configurations: C0, C1, and C4. Among these, the C1 and C4 configurations exhibit the highest switching ratios, while the C0 configuration serves as the fully upward-polarized reference structure. Figure 3 shows the PTS of these three configurations at 300 K. A common feature is that the PTS in the low-frequency region (< 10 THz) is much more significant than in the high-frequency region. At room temperature, low-frequency phonons contribute more significantly to TC[22]. Thus, the differences in PTS in the low-frequency region primarily drive the TC discrepancies among the three configurations. As seen in Fig. 3, the PTS follows a trend of C1 > C0 > C4, indicating the maximum and minimum TCs in C1 and C4 configurations, respectively. This result aligns with the calculated TCs shown in Fig. 2, confirming that TC differences among configurations are mainly due to low-frequency phonons.

To elucidate the mechanism behind the strong polarization-dependent PTS, we calculated the phonon dispersion relations and phonon density of states (DOS) for the C0, C1, and C4 structures using the PHONOPY package[23] considering both strained and unstrained conditions. Figure 4 presents the phonon dispersion relations along the Γ–X path within the 0–12 THz frequency range. Notably, certain structures exhibit phonon band gaps that correlate with TC. For instance, the unstrained C1 structure shows band gaps at 5.5 THz and 7.0 THz, while the strained C1 structure displays a significant band gap at 7.5 THz.

According to Lindsay's phonon-optical band gap theory[24], low-frequency acoustic phonons are primarily responsible for heat conduction, whereas optical phonons provide key scattering channels for acoustic modes through processes such as acoustic–optical–optical (a–o–o) and optical–optical–optical (o–o–o) scatterings, contributing to Umklapp scattering. Phonon scattering theory dictates that the sum of the frequencies of interacting phonons must satisfy energy conservation, often resulting in acoustic and optical phonons combining to produce another optical phonon[22]. As shown in Fig. 4, the C1 structure exhibits the most significant phonon band gaps among the three structures, both with and without strain. Larger phonon band gaps lead to weaker phonon scattering for both a–o–o and o–o–o processes[24]. Consequently, it is expected that the PTS in the low-frequency region—and thus the TC—of the C1 structure is higher than those of the C0 and C4 structures, regardless of strain. This expectation is confirmed by the calculated PTS and TC results shown in Fig. 2 and 3.

In the unstrained structures, the C0 structure possesses a small band gap, whereas the C4 structure has none, implying that phonon scattering is weaker in C0 than in C4. As depicted in Fig. 3, the PTS of C0 is indeed noticeably higher than that of C4, consistent with our analysis. In the strained structures, although the C4 structure also exhibits a band gap, its PTS remains significantly lower than that of the strained C0 structure. An unusually large DOS around 9 THz is observed in the strained C4 structure, indicating the presence of strong localized vibrational modes. These localized modes interact with other phonon modes and impede their propagation, leading to a reduction in PTS and, consequently, a decrease in TC[18]. Therefore, the variations in PTS and TC in polarized o-HfO$_2$ can be attributed to two factors: the influence of phonon band gaps on scattering strength and the degree of phonon localization affecting phonon propagation efficiency.

To confirm this argument, we calculated the phonon participation ratios (PPR) of the different polarization configurations of o-HfO$_2$, which directly correspond to the extent of phonon localization. The PPR quantifies the fraction of atoms participating in a given phonon mode, with values approaching 1/N indicating localized modes and values near 1 indicating delocalized modes[25]. A larger PPR value signifies less phonon localization, while a smaller PPR indicates greater localization[26-28]. Based on previous studies, we adopted PPR < 0.3 as the criterion for identifying predominantly localized phonon modes. Figure 5(a) shows the calculated PPR for the C0 and C4 structures under a 2% strain. In the low-frequency region (2–10 THz), the PPR of the C4 structure is lower than that of C0, indicating more significant phonon localization in C4. This result aligns with the calculated PTS shown in Fig. 3(b), supporting our assertion that localized vibrational modes lead to a reduction in PTS and TC. Furthermore, we note that the general effect of tensile strain on TC for all o-HfO$_2$ structures—namely, that strain typically reduces TC (Fig. 2)—can also be understood from the PPR analysis. Figures 5(b–d) demonstrate that tensile strain leads to a significant reduction in PPR in the C0, C1, and C4 structures within the low-frequency regions (<10 THz). This reduction in PPR corresponds to a pronounced decrease in PTS (as shown in Fig. S3) and, consequently, a decrease in TC (Fig. 2).

# ■ CONCLUSION

In summary, we have proposed a high-efficiency, electrically switchable thermal transistor based on ferroelectric o-HfO$_2$ thin films. Utilizing molecular dynamics simulations with machine learning potentials, we demonstrated that a 24 nm-long o-HfO$_2$ film can achieve four distinct and reversible TC states. These states exhibit a maximum switching ratio of 171% under a 2% tensile strain, significantly surpassing the performance of traditional ferroelectric materials. We uncovered that the strong polarization-dependent TC in o-$HfO_2$ arises from variations in phonon transmission spectra, phonon band structures, and the degree of phonon localization. Specifically, certain polarization configurations introduce larger phonon band gaps, leading to reduced phonon scattering and enhanced TC. Conversely, configurations exhibiting localized vibrational modes result in increased phonon scattering, thereby reducing TC. Our work provides critical insights into the interplay between polarization configuration, strain, and phonon dynamics in o-$HfO_2$, paving the way for the practical realization of next-generation thermal management technologies and advancing the field of thermal transistors towards multi-valued logic computation and neuromorphic computing applications.

## ■ ASSOCIATED CONTENT

**Supplemental Material**

Additional computational details, polarization configuration diagrams not shown in the main text, phonon dispersion relations along the Γ-X-S-Γ-Z high-symmetry path, and comparative analysis of the phonon transmission spectra for the C0 (a), C1 (b), and C4 (c) structures under both strained and unstrained conditions.

## ■ AUTHOR INFORMATION

**Corresponding Authors**

**Zhi-Xin Guo** - *State Key Laboratory for Mechanical Behavior of Materials, Xi'an Jiaotong University, Xi'an, Shaanxi, 710049, China.*
Email: zxguo08@xjtu.edu.cn

**Authors**

**Yong-Kun Huo** - *State Key Laboratory for Mechanical Behavior of Materials, Xi'an Jiaotong University, Xi'an, Shaanxi, 710049, China.*

**Hui-Feng Feng** - *State Key Laboratory for Mechanical Behavior of Materials, Xi'an Jiaotong University, Xi'an, Shaanxi, 710049, China.*

**Chao Yao** - *State Key Laboratory for Mechanical Behavior of Materials, Xi'an Jiaotong University, Xi'an, Shaanxi, 710049, China.*

**Zhongxiao Song** - *State Key Laboratory for Mechanical Behavior of Materials, Xi'an Jiaotong University, Xi'an, Shaanxi, 710049, China.*

Notes

The authors declare no competing financial interest.

# ■ ACKNOWLEDGMENTS

This work was supported by Natural Science Foundation of China (No. 12474237) and Science Fund for Distinguished Young Scholars of Shaanxi Province (No. 2024JC-JCQN-09).

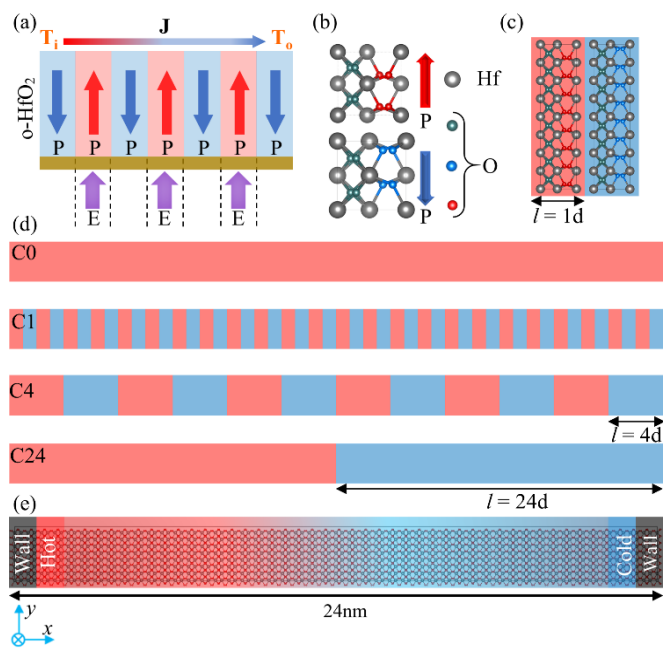

FIG. 1. (a) Schematic diagram for the thermal transistor based on o-HfO$_2$. The brown layer represents the dielectric layer, while the red and blue arrows within o-HfO$_2$ indicate the polarization directions. The purple arrows show the direction of the applied local electric field. (b) Unit cell structure of o-HfO$_2$ in upward and downward polarized states. Silver spheres represent Hf atoms; red and blue spheres denote oxygen atoms in the ferroelectric layer with upward and downward polarization, respectively; green spheres represent oxygen atoms in the spacer layer. (c) Schematic representation of a basic periodic unit (d = 5.04 Å), with red and blue regions corresponding to upward and downward polarization, respectively. (d) Four polarization configurations generated by varying the parameter $l$ shown in (c). (e) Schematic diagram of NEMD simulations for different configurations. The gray region represents the frozen section, while the red and blue regions correspond to the heat and cold baths, respectively.

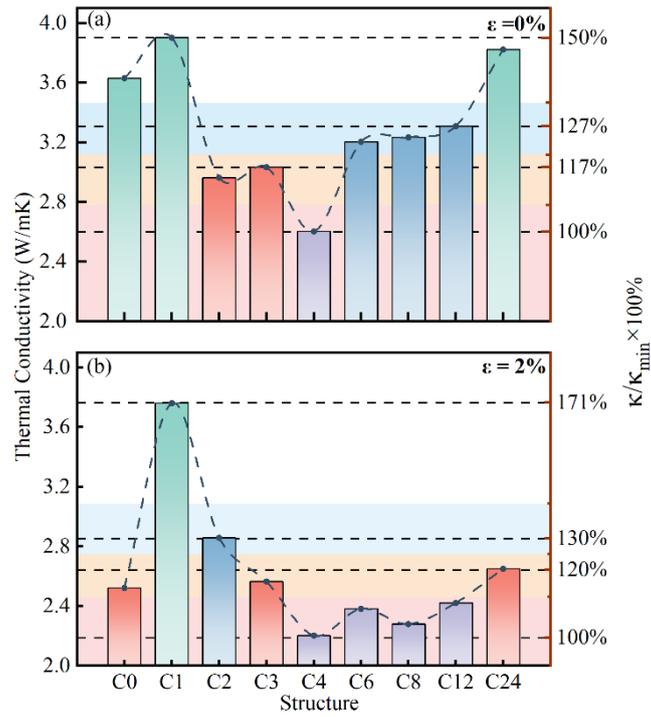

FIG. 2. Thermal conductivities (TC) of various polarization configurations under (a) unstrained ($\varepsilon = 0$) and (b) 2% tensile strain ($\varepsilon = 2\%$) conditions. The right axis in both (a) and (b) represents the ratio of the thermal conductivity for each configuration relative to the minimum thermal conductivity value.

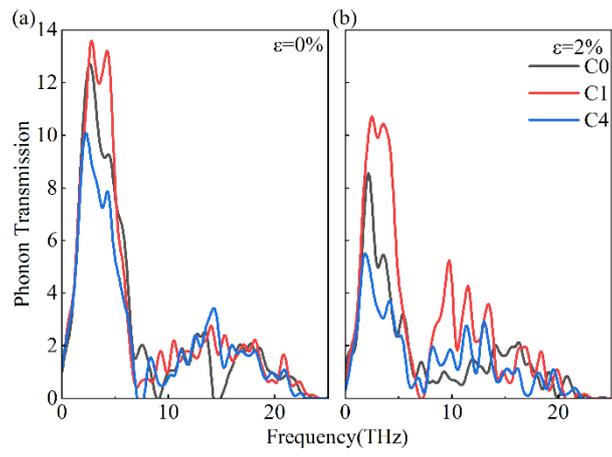

FIG. 3. Phonon transmission spectra (PTS) for the (a) unstrained and (b) 2% tensile strain cases for the C0, C1, and C4 polarization configurations.

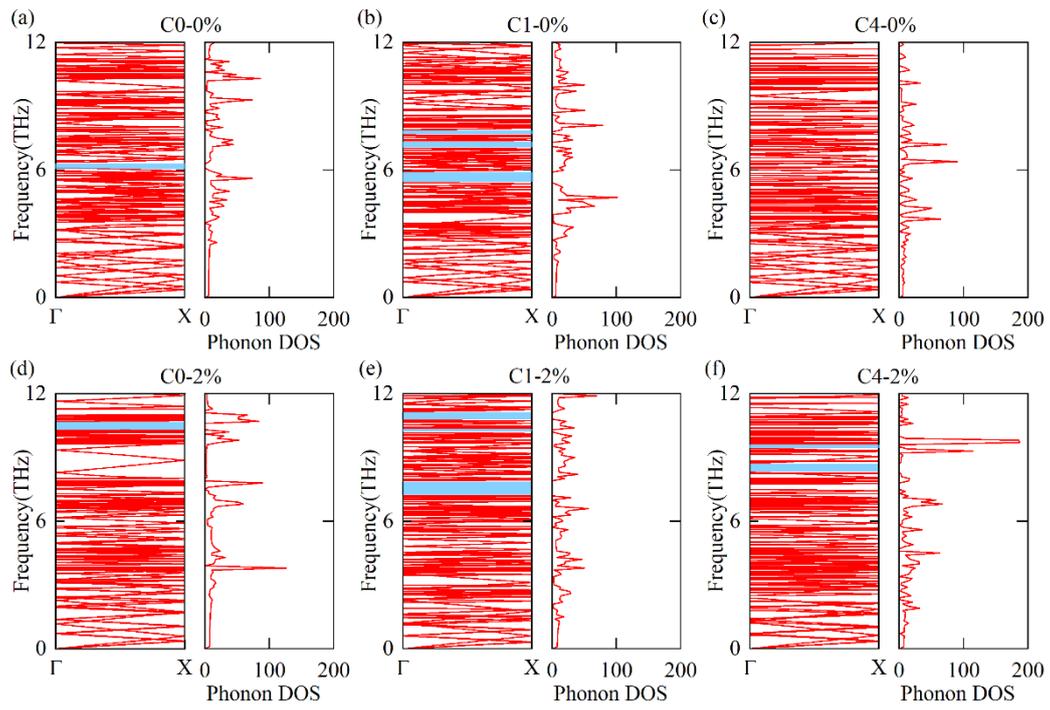

FIG. 4. Phonon dispersion relations along the Γ-X high-symmetry path for the (a) unstrained C0, (b) unstrained C1, (c) unstrained C4, (d) 2% tensile strain C0, (e) 2% tensile strain C1, and (f) 2% tensile strain C4 configurations.

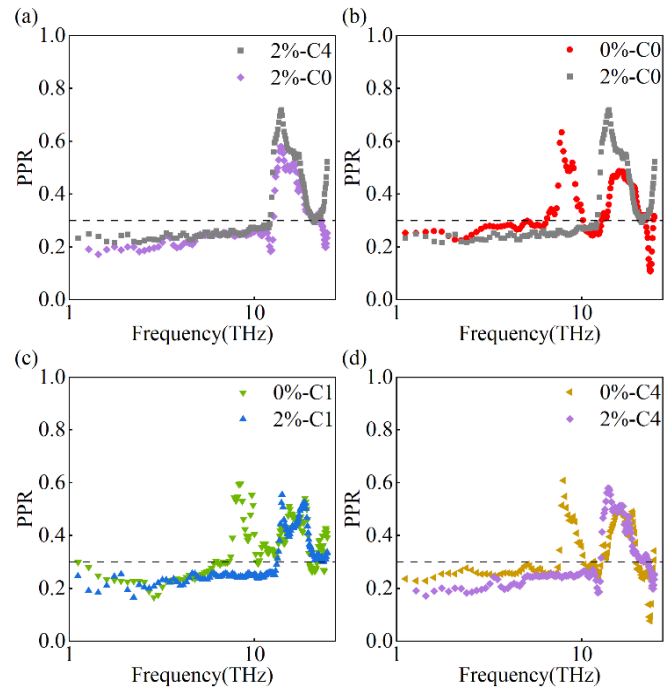

FIG. 5. (a) Phonon participation ratios (PPR) for the C0 and C4 structures under 2% tensile strain. (b), (c), and (d) present comparative PPR analyses of the C0, C1, and C4 structures under strained and unstrained conditions, respectively.